# UNIVERSAL QUANTUM CIRCUIT FOR *n*-QUBIT QUANTUM GATE: A PROGRAMMABLE QUANTUM GATE

PAULO BENÍCIO DE SOUSA and RUBENS VIANA RAMOS

*Departamento de Engenharia de Teleinformática – Universidade Federal do Ceará - DETI/UFC*
*C.P. 6007 – Campus do Pici - 60755-640 Fortaleza-Ce Brasil*

Quantum computation has attracted much attention, among other things, due to its potentialities to solve classical NP problems in polynomial time. For this reason, there has been a growing interest to build a quantum computer. One of the basic steps is to implement the quantum circuit able to realize a given unitary operation. This task has been solved using decomposition of unitary matrices in simpler ones till reach quantum circuits having only single-qubits and CNOTs gates. Usually the goal is to find the minimal quantum circuit able to solve a given problem. In this paper we go in a different direction. We propose a general quantum circuit able to implement any specific quantum circuit by just setting correctly the parameters. In other words, we propose a programmable quantum circuit. This opens the possibility to construct a real quantum computer where several different quantum operations can be realized in the same hardware. The configuration is proposed and its optical implementation is discussed.

## 1. Introduction

Quantum computation has attracted much attention due to its potentialities to solve classical NP problems in polynomial time. The most famous example is the Shor's factorization algorithm [1]. Since Shor's algorithm proposal, much effort has been realized to design and construct quantum computers. Initially, quantum circuits have been proposed to solve some specifics problems, as Deutsch' problem [2], quantum Fourier transform [3,4] and others. The fact that any $2^n \times 2^n$ quantum gate can be built using only single-qubit and CNOT gates [5] encourage the researchers to look for a method to find the quantum circuit that realizes a given quantum gate represented by a unitary $2^n \times 2^n$ matrix. Two decompositions has been used for this task, the cosine-sine decomposition [6] and Cartan's *KAK* decomposition [7,8]. On the other hand, quantum circuit designing using different kinds of genetic algorithm has been successfully reported [9-11]. The main disadvantage of such algorithms is its velocity of convergence for large circuits. However, the result obtained tends to be optimized. Hence, it is possible, nowadays, to find the quantum circuit that realizes any $2^n \times 2^n$ gate, using group theory theorems or artificial intelligence. Usually, the aim is to find the minimal quantum circuit able to solve a given problem. In this work we go in a different direction. Our aim is to propose a general quantum circuit able to implement any specific quantum circuit, by just setting correctly the parameters. In other words, we propose a programmable quantum circuit. This opens the possibility to construct a quantum computer where several different quantum operations can be realized in the same hardware. Since in the universal quantum circuit proposed the parameters are changed setting correctly classical variables as voltage and current, it is possible to use a classical computer to control the universal quantum circuit. This opens the possibility to realize several services for quantum information as quantum control and self-configurable circuits among others. This work is outlined as follow: in Section 2 the general decomposition of single-qubit, two-qubit and three-qubit gates are reviewed. In Section 3 the universal quantum circuit for *n*-qubit quantum gate is presented. In Section 4 a possible optical implementation is discussed. In Section 5 some applications are discussed and, at last, the conclusions are presented in Section 6.

## 2. The general decomposition of single-qubit, two-qubits and three-qubits quantum gates

Any single-qubit gate *U* can be decomposed as:

$$U = e^{i\theta_0} e^{-i\theta_1 \sigma_z/2} e^{-i\theta_2 \sigma_y/2} e^{-i\theta_3 \sigma_z/2} = e^{i\theta_0} \begin{bmatrix} e^{-i\theta_1/2} & 0 \\ 0 & e^{i\theta_1/2} \end{bmatrix} \begin{bmatrix} \cos\left(\frac{\theta_2}{2}\right) & -\sin\left(\frac{\theta_2}{2}\right) \\ \sin\left(\frac{\theta_2}{2}\right) & \cos\left(\frac{\theta_2}{2}\right) \end{bmatrix} \begin{bmatrix} e^{-i\theta_3/2} & 0 \\ 0 & e^{i\theta_3/2} \end{bmatrix} \quad (1)$$

where

$$\sigma_1 = \sigma_x = \begin{bmatrix} 0 & 1 \\ 1 & 0 \end{bmatrix}; \sigma_2 = \sigma_y = \begin{bmatrix} 0 & -i \\ i & 0 \end{bmatrix}; \sigma_3 = \sigma_z = \begin{bmatrix} 1 & 0 \\ 0 & -1 \end{bmatrix} \quad (2)$$

are the Pauli matrices. This decomposition is well known from light polarization theory, since any polarization change can be realized using two retarders and one polarization rotator between them. From (1) we see that a general single-qubit gate is parameterized by three variables.

Any two-qubit gate $U_{AB}$ can be decomposed as [12]:

$$U_{AB} = e^{i\theta_0 I_2 \otimes I_2}(U_A \otimes U_B) U_D (V_A \otimes V_B) \quad (3)$$

$$U_D = \exp\left(-i\sum_{k=1}^{3} \theta_k \sigma_k \otimes \sigma_k\right) = \sum_{n=1}^{4} e^{-i\varphi_n} |\phi_n\rangle\langle\phi_n| \quad (4)$$

$$\varphi_1 = \theta_1 - \theta_2 + \theta_3 \; ; \; \varphi_2 = -\theta_1 + \theta_2 + \theta_3 \; ; \; \varphi_3 = -\theta_1 - \theta_2 - \theta_3 \; ; \; \varphi_4 = \theta_1 + \theta_2 - \theta_3. \quad (5)$$

In (3) $U_A$, $U_B$, $V_A$ and $V_B$ are single-qubit quantum gates while $U_D$ is a non-factorable two-qubit gate responsible for the non-local characteristic of the gate. In (4), $\sigma_k$ are again the Pauli matrices and the quantum states $|\phi_n\rangle$ $n=1,2,3,4$ form the well known magic basis:

$$|\phi_1\rangle = |\Phi^+\rangle = (|00\rangle + |11\rangle)/\sqrt{2} \; ; \; |\phi_2\rangle = -i|\Phi^-\rangle = -i(|00\rangle - |11\rangle)/\sqrt{2} \quad (6)$$

$$|\phi_3\rangle = |\Psi^-\rangle = (|01\rangle - |10\rangle)/\sqrt{2} \; ; \; |\phi_4\rangle = -i|\Psi^+\rangle = -i(|01\rangle + |10\rangle)/\sqrt{2}. \quad (7)$$

In order to find the parameters of the decomposition (3)-(4) let us initially consider the following matrices [7]:

$$M = \frac{1}{\sqrt{2}} \begin{bmatrix} 1 & 0 & 0 & i \\ 0 & i & 1 & 0 \\ 0 & i & -1 & 0 \\ 1 & 0 & 0 & -i \end{bmatrix} \quad (8)$$

$$\Lambda = \begin{bmatrix} 1 & 1 & -1 & 1 \\ 1 & 1 & 1 & -1 \\ 1 & -1 & -1 & -1 \\ 1 & -1 & 1 & 1 \end{bmatrix} \quad (9)$$

Universal quantum circuit for *n*-qubit quantum gate: A programmable quantum gate

If $U$ is the unitary matrix whose decomposition one is looking for, then the following matrices can be defined:

$$U' = M^+ U M \tag{10}$$

$$U_R = \frac{U' + U'^*}{2} \; ; \; U_I = \frac{U' - U'^*}{2i} \tag{11}$$

Now, one can find the generalized singular value decomposition of the matrices $U_R$ and $U_I$:

$$U_R = V_1 C X^+ \tag{12}$$

$$U_I = V_2 S X^+ \tag{13}$$

In (12)-(13), $V_1$, $V_2$ and $X$ are unitary matrices while $C$ and $S$ are nonnegative diagonal matrices. Since $U_I U_R^+$ and $U_I^+ U_R$ are both real and symmetric matrices and, furthermore, $U_R U_R^+ + U_I U_I^+ = U_R^+ U_R + U_I^+ U_I = I$, the unitary matrices $V_1$ and $V_2$ are related by:

$$V_2 = V_1 F \; ; \; V_1 = V_2 F \tag{14}$$

where $F$ is a diagonal matrix whose elements assume only the values +1 and -1. Now, working with (12)-(14) one obtains:

$$U' = U_R + i U_I = V_1 (C + iFS) X^+ \tag{15}$$

$$F = V_1^{-1} V_2 = V_2^{-1} V_1 \tag{16}$$

The matrix $C+iFS$ is a diagonal unitary matrix, hence, it is written as:

$$C + iFS = e^{i\Phi} = \begin{bmatrix} e^{i\Phi_0} & 0 & 0 & 0 \\ 0 & e^{i\Phi_1} & 0 & 0 \\ 0 & 0 & e^{i\Phi_2} & 0 \\ 0 & 0 & 0 & e^{i\Phi_3} \end{bmatrix} \tag{17}$$

and the angles $\theta_k$ in (3)-(4) are then obtained from [7]:

$$\begin{bmatrix} \theta_0 \\ \theta_1 \\ \theta_2 \\ \theta_3 \end{bmatrix} = \Lambda^{-1} \begin{bmatrix} \Phi_0 \\ \Phi_1 \\ \Phi_2 \\ \Phi_3 \end{bmatrix} \tag{18}$$

At last, the local operation $U_A$, $U_B$, $V_A$ and $V_B$ can be obtained from

$$U_A \otimes U_B = MV_1M^+ \tag{19}$$

$$V_A \otimes V_B = MX^+M^+ \tag{20}$$

using some trivial algebra. Hence, a general two-qubit gate is parameterized by fifteen variables (three for the non-local operation and twelve for the locals operations). Using (3)-(4), it was shown in [13] that any two-qubit gate can be constructed using three CNOT gates and eight single-qubit gates, as shown in Fig. 1:

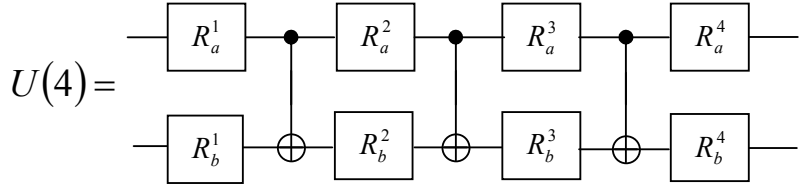

Fig. 1 – Two-qubit gate constructed using eight single-qubit gates and three CNOTs.

The single-qubits gates in Fig. 1 are related to the parameters of (3)-(4) by:

$$R_a^1 = U_A \; ; \; R_B^1 = U_B \tag{21}$$

$$R_a^2 = \frac{i}{\sqrt{2}}(\sigma_x + \sigma_z)e^{-i\left(\theta_1+\frac{\pi}{2}\right)\sigma_x} \; ; \; R_b^2 = e^{-i\theta_3\sigma_z} \tag{22}$$

$$R_a^3 = -\frac{i}{\sqrt{2}}(\sigma_x + \sigma_z) \; ; \; R_b^3 = e^{i\theta_2\sigma_z} \tag{23}$$

$$R_a^4 = V_A\left(\frac{I_2 - i\sigma_x}{\sqrt{2}}\right) \; ; \; R_b^4 = V_B\left(\frac{I_2 - i\sigma_x}{\sqrt{2}}\right)^{-1} \tag{24}$$

where $I_2$ is the 2x2 identity matrix. Hence, given any two-qubit gate, one can use (8)-(20) to find its decomposition and, after, using (21)-(24) to find the equivalent quantum circuit using only single-qubit gates and CNOTs.

Any three-qubit gate $U_{ABC}$ can be decomposed as [14]:

$$U_{ABC} = (A_4 \otimes B_4)N_2(A_3 \otimes B_3)M(A_2 \otimes B_2)N_1(A_1 \otimes B_1) \tag{25}$$

$$N_k = \exp(i(a_k\sigma_x \otimes \sigma_x \otimes \sigma_z + b_k\sigma_y \otimes \sigma_y \otimes \sigma_z + c_k\sigma_z \otimes \sigma_z \otimes \sigma_z)) \tag{26}$$

$$M = \exp(i(a\sigma_x \otimes \sigma_x \otimes \sigma_x + b\sigma_y \otimes \sigma_y \otimes \sigma_x + c\sigma_z \otimes \sigma_z \otimes \sigma_x + dI_2 \otimes I_2 \otimes \sigma_x)) \tag{27}$$

In (25) $A_k$ and $B_k$ ($k$=1,2,3,4) are, respectively, two-qubit and single-qubit gates. The quantum circuits for $N$, $M$ and $U_{ABC}$ are shown in Figs 2, 3 and 4, respectively [14].

Universal quantum circuit for *n*-qubit quantum gate: A programmable quantum gate

$N_{abc}(8) =$

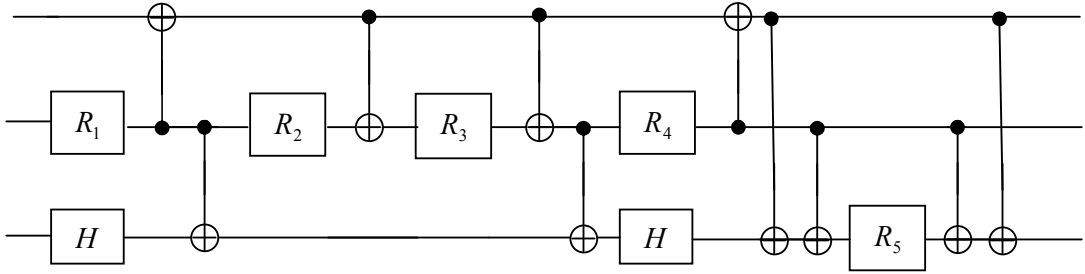

Fig. 2 – Quantum circuit for unitary matrix *N*.

*H* is the Hadamard gate while the other single-qubit gates are:

$$R_1 = e^{-i(-\pi/2)\sigma_z/2} \; ; \; R_2 = e^{-i(2a)\sigma_y/2} \; ; \; R_3 = e^{-i(-2b)\sigma_y/2} \quad (28)$$
$$R_4 = e^{-i(\pi/2)\sigma_z/2} \; ; \; R_5 = e^{-i(2c)\sigma_z/2} \quad (29)$$

$M_{abcd}(8) =$

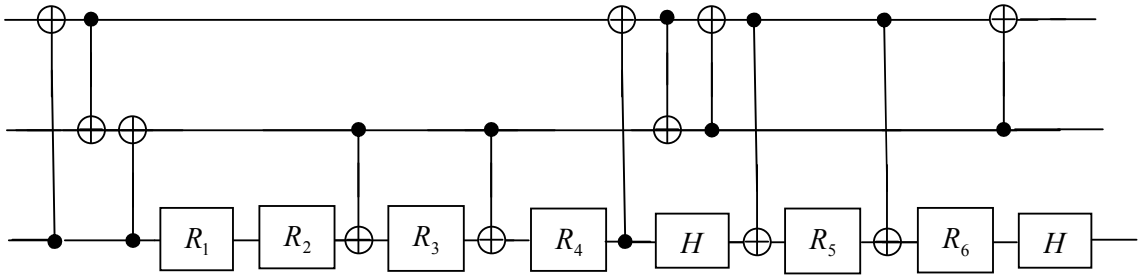

Fig. 3 – Quantum circuit for unitary matrix *M*.

The single-qubit gates are:

$$R_1 = e^{i(\pi/2)\sigma_z/2} \; ; \; R_2 = e^{-i(2a)\sigma_y/2} \; ; \; R_3 = e^{-i(-2b)\sigma_y/2} \quad (30)$$
$$R_4 = e^{-i(\pi/2)\sigma_z/2} \; ; \; R_5 = e^{-i(2c)\sigma_z/2} \; ; \; R_6 = e^{-i(2d)\sigma_z/2} \quad (31)$$

$U_{ABC}(8) =$

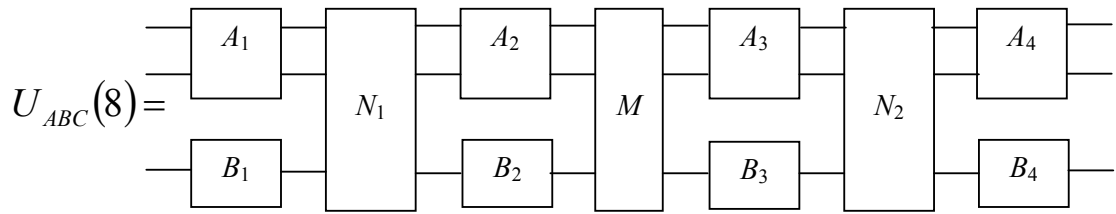

Fig. 4 – Quantum circuit for general three-qubit quantum gate.

Hence, a general three-qubit gate has 82 variables.

## 3. A universal quantum circuit for *n*-qubit quantum gate

Now, we propose a general quantum circuit pattern that can be used to implement any *n*-qubit quantum gate. The idea is to create a basic cell and any quantum gate can be built link the cells. The cell for a general *n*-qubit quantum gate is shown in Fig. 5.

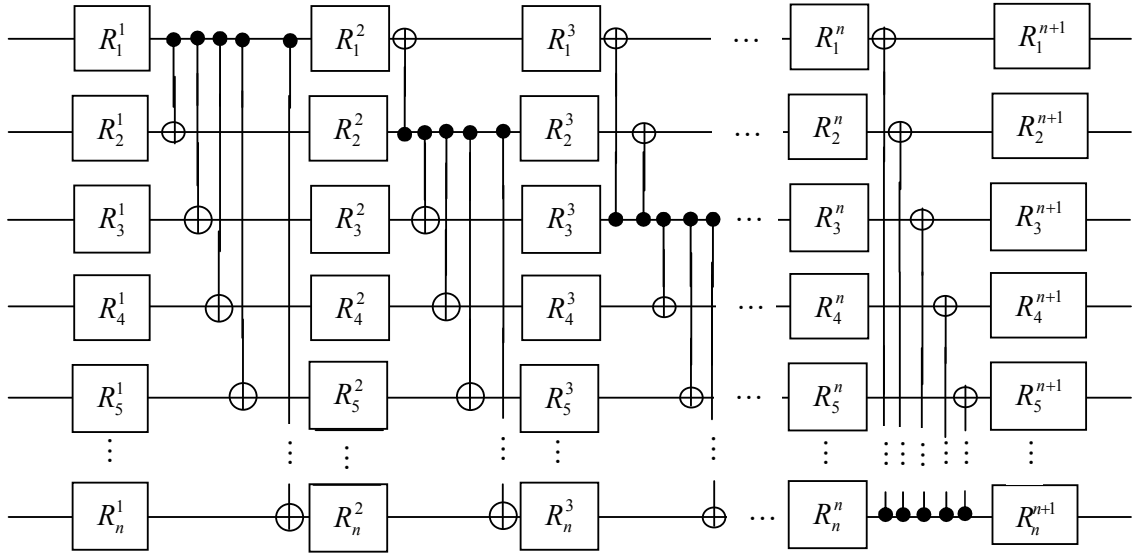

Fig. 5 – Universal cell for any *n*-qubit gate.

As can be seen from Fig. 5, the basic cell is composed by $n(n+1)$ single-qubit gates and $n(n-1)$ CNOTs, hence, any cell has $3n(n+1)+n(n-1)$ variables. For each CNOT there is a variable indicating if it is activated or deactivated (working as unitary gate). Although we have not provided a mathematical proof that any *n*-qubit gate can be built using the scheme of Fig. 5, it is not difficult to notice that this is true. The scheme of Fig. 5 can implement any single-qubit gate in any of the qubits of the quantum bus. It can also implement the CNOT between any two qubits of the quantum bus. Hence, in the simplest case, each cell can represent only a single-qubit gate or a CNOT between any two qubits (all the other gates are set as identity) and several cells can be used to implement any quantum gate. As examples, one can easily observe that any two-qubit gate can be built using two cells of the type:

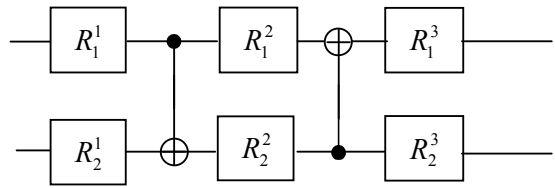

Fig. 6 – Universal cell for two-qubit gates.

while any three-qubit gate can be built using less than 30 sequences of the following cell.

Universal quantum circuit for *n*-qubit quantum gate: A programmable quantum gate

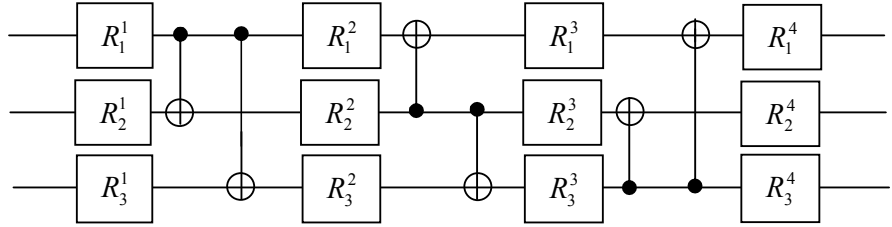

Fig. 7 – Universal cell for three-qubits quantum gates.

Using 7 cells of the type shown in Fig. 7 one can construct the quantum gates *N* and *M* shown in Figs. 2 and 3, respectively. The universal quantum circuit provides the possibility of implementation of a quantum version of the FPGA (Field Programmable Gate Array) technology, which permits the hardware being set via software. In fact, assembling some basic cells for an *n*-qubit quantum circuit, as shown in Fig. 5, several different quantum operations can be implemented just adjusting the parameters of the single-qubit gates and enabling or not the CNOTs gates. When a CNOT gate is not enabled, it will work as an identity gate.

## 4. A possible optical implementation of the programmable quantum gate

In order to implement a programmable quantum gate, one must be able to construct single-qubit gates with easily adjustable parameters and CNOT gates that can be switched to an identity gate. Here, we will use light polarization as the qubit. For this quantum system, as discussed before, any single-qubit gate can be built using two retarders (or compensators) and a polarization rotator between them. Such device can be constructed using optical fibers coils or rotatable wave-plates [15 and references there in]. In this last, any values for retarders and polarization rotator can be achieved rotating the wave-plates (this changes the decomposition of the incident field in the fast and low axes of the birefringent crystals). Hence, in principle, an adjustable single-qubit gate can be implemented using micromechanical adjustment of the rotation of the wave-plates. This method is not a good one because it permits only slow variation of the parameters. However, polarization modulators have been constructed based on bulk III-V semiconductor waveguide microstructures, having potentially small size and high speed modulation [16]. Hence, fully and easily adjustable single-qubit gates for polarization encoded qubits seem to be a problem that can be solved in the near future. A much harder problem is the optical implementation of CNOT gates. A number of solutions have been proposed [17-20]. Here, we will discuss the usefulness of the solution presented in [21-22] in the programmable quantum gate. The set-up proposed in those references, is shown in Fig. 8.

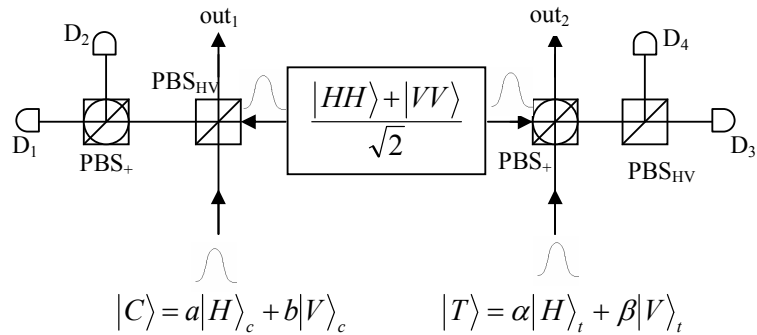

Fig. 8 – CNOT gate implemented with polarization beam splitters (PBS) and two-photon entangled state. $PBS_{HV}$: PBS in horizontal-vertical basis. $PBS_{+}$: PBS in diagonal basis ($\pi/4$, $3\pi/4$). $D_{1-4}$ are single-photon detectors.

The optical setup shown in Fig. 8 implements the CNOT gate only when a single-photon is detected in $D_1$ or $D_2$ and another single-photon is detected in $D_3$ or $D_4$. The output state of the setup in Fig 8 is:

$$|\Psi_f\rangle = \frac{1}{4}\left[|+,H\rangle|\psi_1\rangle + |-,H\rangle|\psi_2\rangle + |+,V\rangle|\psi_3\rangle + |-,V\rangle|\psi_4\rangle\right] + \frac{\sqrt{3}}{2}|\Psi_u\rangle \quad (32)$$

$$|\psi_1\rangle = \left[a\alpha|HH\rangle_{ct} + a\beta|HV\rangle_{ct} + b\alpha|VV\rangle_{ct} + b\beta|VH\rangle_{ct}\right] \quad (33)$$

$$|\psi_2\rangle = \left[-a\alpha|HH\rangle_{ct} - a\beta|HV\rangle_{ct} + b\alpha|VV\rangle_{ct} + b\beta|VH\rangle_{ct}\right] = \left[(XZX) \otimes I\right]|\psi_1\rangle \quad (34)$$

$$|\psi_3\rangle = \left[a\alpha|HV\rangle_{ct} + a\beta|HH\rangle_{ct} + b\alpha|VH\rangle_{ct} + b\beta|VV\rangle_{ct}\right] = \left[I \otimes X\right]|\psi_1\rangle \quad (35)$$

$$|\psi_4\rangle = \left[-a\alpha|HV\rangle_{ct} - a\beta|HH\rangle_{ct} + b\alpha|VH\rangle_{ct} + b\beta|VV\rangle_{ct}\right] = \left[(XZX) \otimes I\right]\left[I \otimes X\right]|\psi_1\rangle \quad (36)$$

In (32) $|\Psi_u\rangle$ is the useless part that contains the situations where none or two photons were detected in the $D_{1\text{-}2}$ and/or $D_{3\text{-}4}$. Further, $|+,H(V)\rangle$ means a single-photon going to $D_1$ and another single-photon going to $D_3$ ($D_4$), while $|-,H(V)\rangle$ means a single-photon going to $D_2$ and another single-photon going to $D_3$ ($D_4$). Observing (32) and (33) one sees that the success probability of CNOT operation is 1/16. However, if one uses single-qubit operations to correct the output state according to where the detections were obtained (detections in $D_2$ and $D_3 \to XZX$ in the control qubit, detections in $D_1$ and $D_4 \to X$ in the target qubit and detections in $D_2$ and $D_4 \to XZX$ in the control qubit and $X$ in the target qubit) the probability of success goes to 1/4.

A crucial component in the CNOT implementation of Fig. 8 is the entangled pair of photons. If instead of $(|HH\rangle+|VV\rangle)/2^{1/2}$ one had used $(|HV\rangle+|VH\rangle)/2^{1/2}$, one would get the unitary operation $(X \otimes I)U_{\text{CNOT}}(X \otimes I)$, that is, a CNOT that inverts the target when the control qubit is $|H\rangle$. On the other hand, if one uses the disentangled state $(|HH\rangle+|VH\rangle)/2^{1/2}$ instead of the Bell state, the setup in Fig. 8 implements the identity operation. Hence, the programmable universal gate can be constructed using controlled CNOTs as shown in Fig. 8, but having the possibility to choose from $(|HH\rangle+|VV\rangle)/2^{1/2}$ and $(|HH\rangle+|VH\rangle)/2^{1/2}$. If the CNOT has to be activated, the two-photon state $(|HH\rangle+|VV\rangle)/2^{1/2}$ is used, otherwise, the state $(|HH\rangle+|VH\rangle)/2^{1/2}$ is used.

The main problems with CNOT implementation of Fig. 8 are the necessity of reliable entangled two-photon sources and its probabilistic behavior. This last is the price to be paid for trying to implement a non-linear operation using linear devices. However, it has been shown [23] that the non-linearity of quantum measurements can be used to implement (almost) deterministic quantum gates. A near determinist CNOT using quantum non-demolition measurement has been proposed in [24].

## 5. Applications of the programmable quantum gate

The first application is the construction of the quantum prisoner's dilemma game [254-27], for two players, using two cells of the universal quantum circuit for two-qubit gate. The main objective of a quantum game, in a non-cooperative game, is to achieve an optimal equilibrium (Pareto) in situations where the classical strategies can obtain only the Nash's equilibrium (where both players are not encouraged to change their decisions). The main idea is to gain advantage over "a classical opponent", by using of quantum strategies. In fact, when the classical scenario is considered, a player can evaluate only one alternative and take one single decision (normally irreversible). But, when the intrinsic parallelism inside quantum strategies (using the superposition of quantum states) is used, the player can consider multiple alternatives simultaneously.

Many of these desired properties can be observed using the prisoner dilemma, that is described in short as follows: if we consider that two people are been accused to commit some crime and there is no possibility to realize any communication between them, which characterizes a non-cooperative game,

Universal quantum circuit for *n*-qubit quantum gate: A programmable quantum gate they have two possible strategies: cooperate (C) or defect (D). The following table shows the possible actions and their payoffs for both players [25-27 and references there in].

| Alice \ Bob | C | D |
|---|---|---|
| C | (3,3) | (0,5) |
| D | (5,0) | (1,1) |

Table 2 – Table of payoffs for prisioner's dilemma.

In the classical scenario, each player is not supposed to know that the adversary can cooperate (which implies in a risk), so the common decision is to defect. This situation – the Nash equilibrium – is important because there is no incentive for changing this decision without running a risk. Hence, there is no rational assumption in which they can get an optimal final situation – the Pareto optimal – in which both would cooperate and, together, they could reach the best situation.

Now, considering the quantum scenario, the players' decisions are implemented through quantum gates. The strategies to cooperate and defect are represented by the quantum states $|C\rangle$ and $|D\rangle$, respectively. The initial total state is assumed to be the tensor product $|CC\rangle$. The total state is processed by a quantum gate $J$ that entangles both qubits, $J|CC\rangle=\cos(\gamma/2)|CC\rangle+\sin(\gamma/2)|DD\rangle$. The kind of solutions obtained depends on the amount of entanglement created [26]. The strategies are implemented, by each player individually, applying a single-qubit quantum gate. The setup is illustrated in Fig. 9.

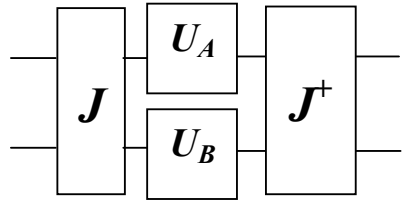

Fig. 9 – Quantum circuit for quantum prisioners dilemma.

The quantum circuit of Fig. 9 can be easily implemented using two cells of the universal quantum circuit for two-qubit gate, as shown in Fig. 10.

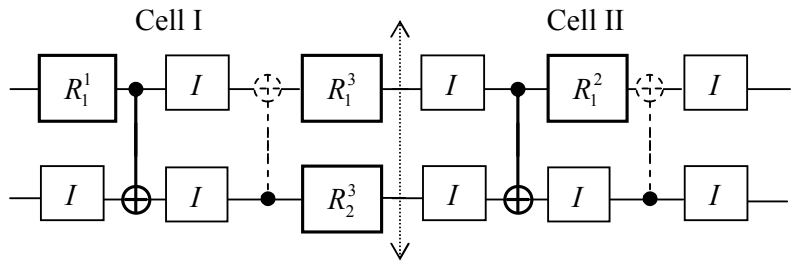

Fig. 10 – Quantum prisoner's dilemma implementation using cells of the universal quantum circuit for two-qubit gate.

In Fig. 10, for cell I, $R_1^1 = \exp(i\gamma\sigma_y/2)$, $R_1^3 = U_A$ and $R_2^3 = U_B$ while for cell II, $R_1^2 = (R_1^1)^{-1}$. Hence, the game can be played changing $\gamma$, that controls the amount of entanglement created, and changing the strategies through of the variation of $U_A$ and $U_B$.

Now, using two cells of the universal quantum circuit for three-qubit gate, one can construct, for example, a Toffoli gate, as shown in Fig. 11

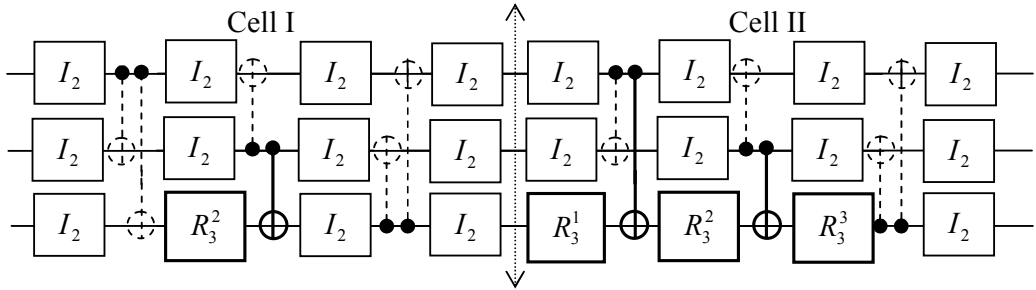

Fig. 11 – Toffoli gate constructed from two cells of the universal quantum circuit for three-qubit gate.

For Cell I $R_3^2 = \exp(i\pi/4\sigma_y)$ and for Cell II $R_3^1 = \exp(i\pi/4\sigma_y)$, $R_3^2 = R_3^3 = \exp(-i\pi/4\sigma_y)$. On the other hand, the Toffolli gate operated in level 0 instead of 1 is constructed as shown in Fig. 12.

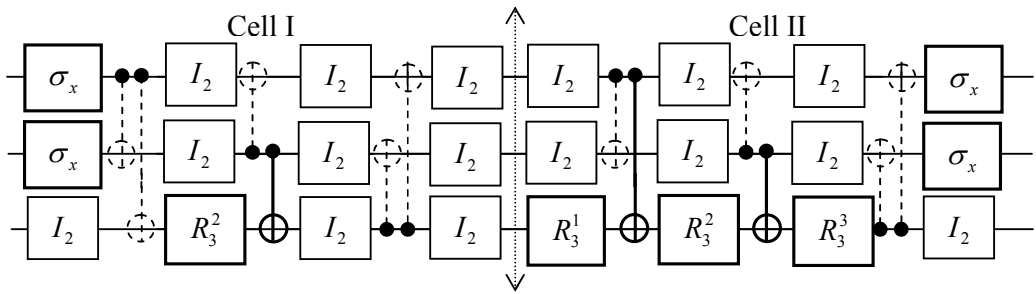

Fig. 12– Toffoli gate, operating in the level 0, constructed from two universal cells for three-qubits quantum gates.

Hence, one can change from a Toffoli gate operating in level 1 to a Toffoli gate operating in level 0 by just changing the parameters of 4 single-qubit gates.

## 6. Conclusions

We proposed a programmable quantum circuit, that is, a general quantum circuit, based on single-qubit gates and CNOTs, able to implement any *n*-qubit quantum gate, by just setting the parameters correctly. Such parameters, for optical single-qubit gates, are the compensators and polarization rotator parameters, that is, the parameters of a polarization modulator, which can be constructed using movable wave-plates controlled micromechanically or semiconductors waveguides. In the case of CNOTs, the parameter is a signal control that activates or not the CNOT. If the CNOT is deactivated it works as an identity matrix. An example using optical CNOT gate was presented. At last, some examples of the use of the universal quantum circuit proposed were presented: It was shown the implementation of a quantum game of two players (the prisoner's dillema) and the implementation of the Toffoli gate operated in level 1 and 0.

## Acknowledgements

This work was supported by the Brazilian agency FUNCAP.

Universal quantum circuit for *n*-qubit quantum gate: A programmable quantum gate